\documentclass[prd,twocolumn,showpacs]{revtex4}
\usepackage{graphicx,amsmath,amssymb}
\newcommand{\haak}[1]{\left(#1\right)}
\newcommand{\rhaak}[1]{\left [#1\right]}
\newcommand{\lhaak}[1]{\left | #1\right |}

\newcommand{\gem}[1]{\left\langle #1\right\rangle}

\newcommand{\half}{\frac{1}{2}}

\newcommand{\arctanh}{\operatorname{arctanh}} 
\renewcommand{\imath}{\text{i}}

\begin{document}

\title{Uranus's anomalously low excess heat constrains strongly interacting dark matter} 
\author{Saibal Mitra}

\affiliation{Instituut voor Theoretische Fysica,\\ Valckenierstraat 65,\\ 1018 XE Amsterdam,\\ The Netherlands}
\pacs{95.35.+d}
\date{\today}
\begin{abstract}
Dark matter annihilations can generate significant amounts of internal heat inside planets if dark matter consists mainly of particles with nuclear cross sections in the micro-barn range or larger (SIMPs). By considering a detailed model of Uranus's interior, we calculate upper limits on the S-wave annihilation cross section for these particles as a function of their mass. These upper limits, together with other experimental and theoretical constraints, rule out SIMPs with masses between 150 MeV and $10^{4}$ GeV.
\end{abstract}

\maketitle

\section{Introduction}
Besides weakly interacting massive particles, many other candidates have been proposed for dark matter. See \cite{mun} for a comprehensive review. Strongly Interacting Massive Particles (SIMPs) have been proposed as a solution for the ultra high energy cosmic ray problem \cite{chung} and for the absence of cooling flows \cite{chuz}. Constraints from experiments \cite{xqc,mcg} and from Big Bang nucleosynthesis and gamma ray observations \cite{cyb}, limit SIMPs with cross sections on protons in the micro-barn range to be lighter than 1 GeV or to be heavier than $10^{4}$ GeV. See \cite{zah} for accurate calculations of the allowed window below 4 GeV. We note that, despite the Lee-Weinberg bound \cite{leewein}, dark matter can consist of particles lighter than 1 GeV \cite{boehm,zah}.

In this article we show that annihilations of SIMPs can generate significant amounts of heat inside planets. Obviously, the heat production due to annihilations inside the planet has to be smaller than the total internal heat production. This provides us with a new way to constrain SIMPs. Large planets with very low excess heat pose the strongest constraints. Uranus's internal heat production is atypically small, only about a tenth of the similar sized planet Neptune \cite{pearl}. In \cite{far} this idea was used to exclude the H dibaryon model and in \cite{new} new constraints have been obtained for other baryonic dark matter candidates.

The capture rate of SIMPs is negligible unless $n\sigma R_{\text{planet}}>>1$, where $n$ is the density of nucleons in the planet, $R_{\text{planet}}$ the radius of the planet and $\sigma$ the average elastic cross section. This means that $\sigma$ needs to be larger than about $10^{-32}$ cm$^2$ for SIMPs to generate a significant amount of heat inside large planets such as Uranus. We can crudely estimate an upper bound on the heat production, $W_{\text{max}}$, by assuming that the entire dark matter flux is transferred into heat.
Assuming a relic dark matter density of 0.3 GeV/cm$^3$ and a typical velocity of 270 km/s of a dark matter particle w.r.t.\ the solar system we get:
\begin{equation}\label{crude}
\begin{split}
W_{\text{max}} & \approx \pi R_{\text{planet}}^2 \times \haak{0.3 \text{ GeV}\text{ cm}^{-3}}\times \haak{270 \text{ km/s}}\\ & \approx 4 \haak{\frac{R_{\text{planet}}}{10^4 \text{ km}}}^{2}\times 10^{15} \text{ W}
\end{split}
\end{equation}
In case of Uranus, $R_{\text{planet}}=2.509\times 10^4$ km and thus $W_{\text{max}}\approx 2.5\times 10^{16}$ W, which is much larger than the observed heat production of $\haak{3.4\pm 3.8}\times 10^{14}$ W \cite{pearl}.

To calculate the actual heat generated by dark matter annihilations in a SIMP dark matter scenario, we have to take into account that, because of the mass mismatch with nuclei, SIMPs need many collisions to lose enough energy to become captured inside a planet. For low mass SIMPs this means that a significant part of the incident flux of SIMPs will end up being reflected back into space. Also, if the annihilation cross section is low enough, most SIMPs will eventually evaporate rather than annihilate. The evaporation rate depends not only on the mass of the SIMPs but also on the elastic cross section. Particles escaping from the gravity of a planet typically escape from one scattering length below the surface. Particles with lower elastic cross sections will thus escape from deeper below the surface, where it is hotter. This means that these particles will have higher escape rates.

We will use a detailed model of Uranus's interior to take these effects into account for SIMPs lighter than 1 GeV interacting with spin independent interactions with nuclei. Upper limits on the S-wave annihilation cross section will be derived as a function of the mass and the elastic cross section on protons.

\section{Capture rate of light SIMPs}
The velocity distribution, $f\haak{v}$, of halo dark matter particles is usually assumed to be Maxwellian \cite{bin}:
\begin{equation}
f\haak{v}=\haak{\pi v_{0}^{2}}^{-\frac{3}{2}}\exp\rhaak{-\haak{\frac{v}{v_{0}}}^{2}}.
\end{equation}
Here $v_{0}\approx 220 \text{ km/s}$ \cite{chr} is the local circular speed. This velocity distribution has to be cut off at the escape speed of the galaxy. Since particles in the tail of the velocity distribution will make only a small contribution to the captured flux, we will ignore this. The solar system is moving around the center of the galaxy with a speed of about $v_{0}$. This means that the velocity distribution w.r.t.\ the solar system is anisotropic and rather complicated. However, since we have to integrate the flux of the SIMPs over the surface of a planet, we can replace the actual velocity distribution by the one averaged over all directions. This averaged velocity distribution, $\tilde{f}\haak{v}$, is:
\begin{equation}
\begin{split}
\tilde{f}\haak{v} & = \frac{1}{4 \pi v_{0}^{2}}\int_{\lhaak{u}=v_{0}}d^{2}u f\haak{v-u}\\
& = \half e^{-1}
\haak{\pi v_{0}^2}^{-\frac{3}{2}}\exp\rhaak{-\haak{\frac{v}{v_{0}}}^2}\frac{v_{0}}{\lhaak{v}}\sinh\haak{\frac{2 \lhaak{v}}{v_{0}}}.
\end{split}
\end{equation}

A SIMP entering a planet will typically undergo multiple scatterings before it is either reflected back into space or is thermalized and captured. In \cite{zah} this problem is investigated for the Earth for dark matter particles heavier than 2 GeV using simulations. In the Appendix we fit these results to an approximate expression obtained by treating this problem in the continuum limit, where it becomes a diffusion problem. We find that the capture rate, denoted as $\mathcal{R}_{\text{capt}}$, is:
\begin{equation}\label{rcapt}
\mathcal{R}_{\text{capt}}\approx 4\pi R_{\text{planet}}^{2}n_{\text{simp}}\int_{v_{\text{esc}}}^{\infty} R\haak{v}\pi\tilde{f}\haak{v}v^3 dv
\end{equation}
where $n_{\text{simp}}\approx 0.3\text{ GeV}/(m_{\text{simp}}\text{cm}^3)$ is the halo number density of SIMPs, $v_{\text{esc}}$ is the escape velocity from the planet and $R\haak{v}$ is the probability that a SIMP moving into the planet with speed $v$ is captured:
\begin{equation}\label{capprob}
R\haak{v}\approx 1.27\times \rhaak{\log{\haak{\frac{v}{v_{\text{esc}}}}}}^{-\frac{1}{2}}\sqrt{\frac{m_{\text{simp}}}{m_{\text{eff}}}}.
\end{equation}
Here $m_{\text{eff}}$ is an effective nucleon mass, defined as:
\begin{equation}\label{meff}
\haak{m_{\text{eff}}}^{-1}\equiv\sum_{j}f_{j}m_{j}^{-1}
\end{equation}
where $m_{j}$ denotes the mass of a nucleus of type $j$ and $f_{j}$ is the relative probability that a SIMP will scatter off a nucleus of this type. $f_{j}$ is thus proportional to $n_{j}\sigma_{j}$, where $n_{j}$ is the number density of nuclei of type $j$ and $\sigma_{j}$ is the elastic cross section with nuclei of that type. For spin independent interactions the $\sigma_{j}$ can be related to the elastic cross section with protons as follows \cite{lew}:
\begin{equation}
\sigma_{j}=\haak{\frac{\mu_{j}}{\mu_{\text{prot}}}}^{2}A_{j}^2\sigma_{\text{prot}}.
\end{equation}
Here $\sigma_{\text{prot}}$ is the cross section with protons, $A_{j}$ is the number of nucleons in nuclei of type $j$, $\mu_{j}$ is the relative mass of a SIMP and a nucleus of type $j$ and $\mu_{\text{prot}}$ is the relative mass of a SIMP and a proton. We will assume that SIMPs undergo most of their scatterings before being captured in the upper 20\% of Uranus's atmosphere. There 93 \% of the nucleons are protons and about 7\% are helium nuclei. According to \eqref{meff}, the effective nuclear mass is:
\begin{equation}
m_{\text{eff}}=\frac{0.93\haak{1+\frac{m_{\text{simp}}}{4 m_{\text{prot}}}}^2+1.12\haak{1+\frac{m_{\text{simp}}}{ m_{\text{prot}}}}^2}{0.93\haak{1+\frac{m_{\text{simp}}}{4 m_{\text{prot}}}}^2+0.28\haak{1+\frac{m_{\text{simp}}}{m_{\text{prot}}}}^2}.
\end{equation}
And the average cross section, $\sigma$, is:
\begin{equation}
\sigma=\rhaak{0.93+1.12\haak{\frac{1+\frac{m_{\text{simp}}}{m_{\text{prot}}}}
{1+\frac{m_{\text{simp}}}{4 m_{\text{prot}}}}}^2}\sigma_{\text{prot}}.
\end{equation}

\section{Evaporation and annihilation of SIMPs}
To estimate the evaporation and annihilation rates of captured SIMPs, we need to know the density profile of the SIMPs. As long as temperatures don't change too rapidly over the course of the length of one mean free path of a SIMP, we can assume that the SIMPs are in local thermal equilibrium with the nuclei in Uranus's interior. Local thermal equilibrium will break down high in Uranus's atmosphere where densities are low. Since only a small fraction of the SIMPs is expected to reside there, this can be neglected in a first approximation.

\begin{figure}[h]
\begin{center}
\includegraphics[width=0.5\textwidth]{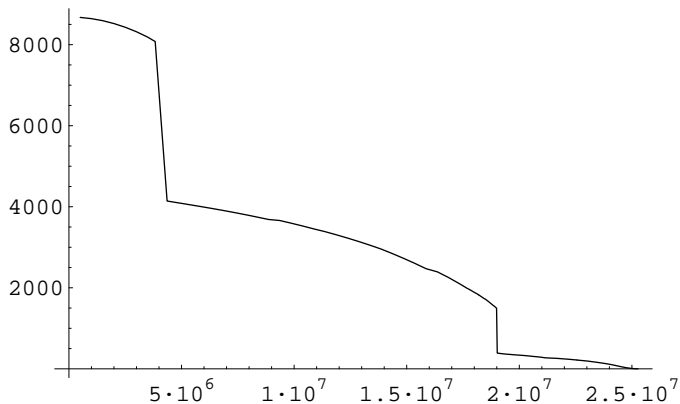}
\caption{Density inside Uranus in kg/m$^3$ as a function of the radius in meters.}\label{dens}
\end{center}
\end{figure}

\begin{figure}[h]
\begin{center}
\includegraphics[width=0.5\textwidth]{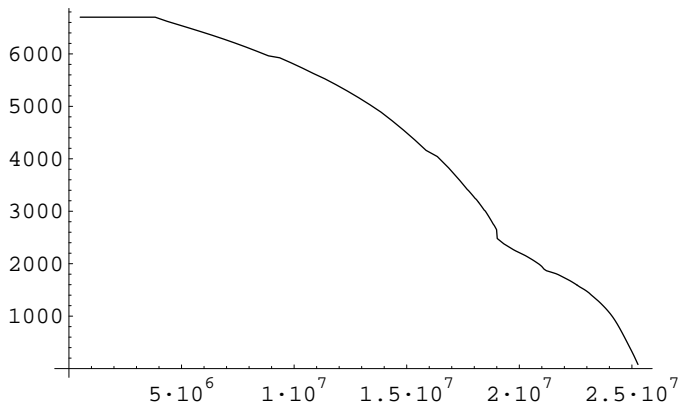}
\caption{Interior temperature (K) as a function of the radius in meters.}\label{temp}
\end{center}
\end{figure}

\begin{figure}[h]
\begin{center}
\includegraphics[width=0.5\textwidth]{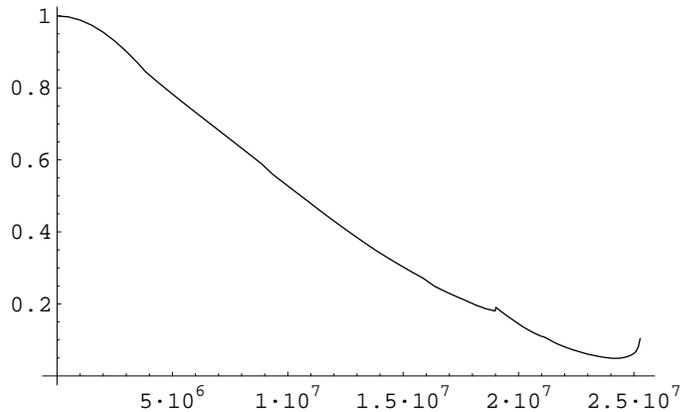}
\caption{Density profile of 0.5 GeV SIMPs in Uranus, normalized to 1 at the center, as a function of the radius in meters.}\label{prof}
\end{center}
\end{figure}

The density profile, $n_{\text{simp}}\haak{r}$, follows from demanding that the SIMPs are in hydrostatic equilibrium:
\begin{equation}\label{nldm}
n_{\text{simp}}\haak{r}=\frac{n_{\text{simp}}\haak{0}T\haak{0}}{T\haak{r}}\exp\rhaak{-\frac{m_{\text{simp}}G}{k}\int_{0}^{r}\frac{M\haak{y}}{y^2 T\haak{y}}dy}.
\end{equation}
Here $G$ is the gravitational constant and $M\haak{y}$ is the cumulative mass of Uranus up to radius $y$. Even though it is impossible to measure the density and temperature profile of Uranus, reasonable accurate models can be made based on reasonable assumptions about the interior composition, the measured heat flux, the total mass and gravitational moments \cite{pod}. The density profile and temperature profiles based on such a model are shown in Fig.\ \ref{dens} and Fig.\ \ref{temp} respectively \footnote{I am very grateful to Morris Podolak for generating the temperature and density profiles for the purpose of this article.}. In Fig.\ \ref{prof} the density profile of 0.5 GeV SIMPs is shown.


The evaporation rate, $\mathcal{R}_{\text{evap}}$, can be estimated by calculating the upward flux of SIMPs from one scattering length below the surface \footnote{This is essentially the same approximation that yields Jean's escape formula for planetary atmospheres.}:
\begin{equation}\label{revap}
\begin{split}
\mathcal{R}_{\text{evap}}\approx &\sqrt{\frac{k T\haak{r_*}}{2\pi m_{\text{simp}}}}\exp\haak{-\frac{m_{\text{simp}} v_{\text{esc}}\haak{r_*}^{2}}{2 k T\haak{r_*}}}\\
&\times\haak{1+\frac{m_{\text{simp}} v_{\text{esc}}\haak{r_*}^{2}}{2 k T\haak{r_*}}}4\pi r_*^{2}n_{\text{simp}}\haak{r_*}.
\end{split}
\end{equation}
Here $r_{*}$ is:
\begin{equation}
\int_{r_{*}}^{\infty}n\haak{r}\sigma dr = 1
\end{equation}
and $v_{\text{esc}}\haak{r_{*}}$ is the escape velocity at $r=r_{*}$.

The annihilation rate, $\mathcal{R}_{\text{ann}}$, can be expressed as:
\begin{equation}\label{rann}
\begin{split}
\mathcal{R}_{\text{ann}}&=\int d^{3}x n_{\text{simp}}\haak{x}^{2}\gem{\sigma_{\text{ann}}v}\\
&=a\int d^{3}x n_{\text{simp}}\haak{x}^2+\frac{6 k}{m c^2} b\int d^{3}x n_{\text{simp}}\haak{x}^2 T\haak{x}
\end{split}
\end{equation}
Where we have assumed that $\sigma_{\text{ann}}$ is well approximated by the S-wave and P-wave contributions:
\begin{equation}
\sigma_{\text{ann}}v=a+b\frac{v^{2}}{c^{2}}.
\end{equation}

\section{Constraints on $\sigma_{\text{ann}}$ }
Besides satisfying the constraints from Uranus's excess heat, the annihilation cross section has to be consistent with the SIMPs being thermal relics. According to \cite{boehm} for particles in the MeV-GeV range:
\begin{equation}
10^{-27} \text{ cm}^{3}\text{/s}\lesssim\gem{\sigma_{\text{ann}} v}\lesssim 10^{-26}\text{ cm}^{3}\text{/s}
\end{equation}
during the freeze-out epoch. For $m_{\text{simp}}$ in the range of a few hundred MeV to 1 GeV the freeze-out temperature is about $m_{\text{simp}}c^{2}/\haak{20 k}$. This means that at freeze-out $\sigma_{\text{ann}}v\approx a+b/3$. The lower limit for the annihilation cross section during freeze-out thus implies that:
\begin{equation}\label{therm}
a+b/3 \gtrsim 10^{-27} \text{ cm}^{3}\text{/s}.
\end{equation}

Uranus's excess heat, as measured by Voyager 2 during its 1986 fly-by, is $\haak{3.4\pm 3.8}\times 10^{14}$ W \cite{pearl}. Models of Uranus's atmosphere based on the observed temperature profile in Uranus's upper atmosphere suggest that Uranus's internal heat is close to Voyager's 1-$\sigma$ upper bound \cite{marl}. We shall consider SIMP models that predict that more than $10^{15}$ W of energy is generated by annihilations as excluded. Although only part of the energy generated by annihilations can contribute to the internal heat, it is unlikely that a significant fraction of Uranus's internal heat is produced by exotic processes such as annihilations of SIMPs.

To estimate the energy produced by SIMP annihilations, consider the time evolution of the total number of captured SIMPs, $\mathcal{N}\haak{t}$:
\begin{equation}\label{ntr}
\frac{d\mathcal{N}\haak{t}}{dt}=\mathcal{R}_{\text{capt}}-\mathcal{R}_{\text{evap}}-\mathcal{R}_{\text{ann}}.
\end{equation}
It follows from \eqref{nldm}, \eqref{revap} and \eqref{rann} that there exists functions $\alpha\haak{m_{\text{simp}},\sigma_{\text{prot}}}$ and $\beta\haak{m_{\text{simp}}}$ such that:
\begin{eqnarray}
\mathcal{R}_{\text{evap}}&=&\alpha\haak{m_{\text{simp}},\sigma_{\text{prot}}} \mathcal{N},\\\label{pwr}
\mathcal{R}_{\text{ann}}&=& \beta\haak{m_{\text{simp}}}\mathcal{N}^{2}.
\end{eqnarray}
Note that $\beta\haak{m_{\text{simp}}}$ is a linear combination of the parameters $a$ and $b$. We set $t=0$ at the time of formation of the solar system, about 4.5 billion years ago. Thus $\mathcal{N}\haak{0}=0$ and \eqref{ntr} implies that:
\begin{equation}
\begin{split}
\mathcal{N}\haak{t}=& \sqrt{\frac{\alpha^{2}}{4\beta^{2}}+\frac{\mathcal{R}_{\text{capt}}}{\beta}}\tanh\haak{\half\sqrt{\alpha^{2}+4\beta\mathcal{R}_{\text{capt}}}\haak{t+C}}\\
&\mbox{}-\frac{\alpha}{2\beta},
\end{split}
\end{equation}
where
\begin{equation}
C=\frac{2}{\sqrt{\alpha^{2}+4\beta\mathcal{R}_{\text{capt}}}}\arctanh\haak{\frac{\alpha}{\sqrt{\alpha^{2}+4\beta\mathcal{R}_{\text{capt}}}}}.
\end{equation}
The total energy generated by SIMP annihilations now follows by substituting this in \eqref{pwr} and putting $t=4.5$ billion years.

Demanding that $m_{\text{simp}}c^{2}\beta\mathcal{N}\haak{t}^{2}<10^{15}$ W yields the following constraints on the annihilation cross section. For every $m_{\text{simp}}$ there exists a critical value, $\sigma_{\text{crit}}$ for $\sigma_{\text{prot}}$. If $\sigma_{\text{prot}}$ is smaller than $\sigma_{\text{crit}}$, there are no constraints on $a$ and $b$ except for the usual thermal relic constraints. If $\sigma_{\text{prot}}$ is larger than $\sigma_{\text{crit}}$, $a$ has to be smaller than $10^{-27}$ cm$^3$/s. According to \eqref{therm} this means that $b$ can no longer be zero. The upper limit for $a$ quickly reaches a terminal value, $a_{\text{max}}$, for $\sigma_{\text{prot}}$ larger than $\sigma_{\text{crit}}$. To a good approximation this happens for $\sigma_{\text{prot}}\gtrsim 3 \sigma_{\text{crit}}$. The values for $\sigma_{\text{crit}}$ and $a_{\text{max}}$ for $m_{\text{simp}}$ in the range 0.8 GeV to 0.3 GeV are shown in Table \ref{limits}. 

\begin{table}[h]
\begin{center}
\begin{tabular}{|l|l|l|}\hline
$\frac{m_{\text{simp}}}{\text{GeV}}$ & $\frac{\sigma_{\text{crit}}}{10^{-31}\text{cm}^{2}}$ & $\frac{a_{\text{max}}}{10^{-34}\text{cm}^{3}/\text{s}}$\\\hline
0.8&1.6&9.0  \\\hline
0.7&2.9&7.0  \\\hline
0.6&5.6&8.8  \\\hline
0.5&12& 11 \\\hline
0.4&28& 15 \\\hline
0.3&86& 17 \\\hline

\end{tabular}
\caption{Constraints on the annihilation cross section as a function of $m_{\text{simp}}$ and $\sigma_{\text{prot}}$. If $\sigma_{\text{prot}}>\sigma_{\text{crit}}$, $a$ has to be smaller than $10^{-27}$ cm$^{3}$/s and thus $b$ can no longer be zero. If $\sigma_{\text{prot}}\gtrsim 3 \sigma_{\text{crit}}$, $a$ has to be smaller than $a_{\text{max}}$.}\label{limits}
\end{center}
\end{table}

Since it is to be expected that the S-wave cross section should not differ from the elastic cross section by more than a few orders of magnitude, these results rule out SIMPs with cross sections larger than $\sigma_{\text{crit}}$.

For $m_{\text{simp}}<0.3$ GeV, the density of SIMPS becomes large at the outer regions of Uranus's atmosphere. To determine $\sigma_{\text{crit}}$ and $a_{\text{max}}$ with reasonable accuracy would require using detailed models of Uranus's upper atmosphere. We estimate that for $m_{\text{simp}}< 0.15$ GeV, SIMPs cannot be constrained and that at $m_{\text{simp}}= 0.15$ GeV, $\sigma_{\text{prot}}$ has to be smaller than  $10^{-27}$ cm$^{2}$. For $m_{\text{simp}}>0.8$ GeV, $\sigma_{\text{crit}}$ becomes so low that local thermal equilibrium starts to break down. It is clear, however, that SIMPs in the mass range 0.8 - 1 GeV are excluded as well. Above 1 GeV known experimental constraints exclude SIMPs up to masses of about $10^{4}$ GeV \cite{xqc,mcg}.

\section{Conclusions and outlook}
We have shown that Uranus's low excess heat rules out SIMPs as dark matter candidates for $150\text{ MeV}\lesssim m_{\text{simp}}<1$ GeV. SIMPs with masses in the range 1-$10^{4}$ GeV were already ruled out \cite{mcg}. It would be interesting to constrain SIMPs above this mass range. Preliminary results indicate that for $10^{4}\text{ GeV}\lesssim m_{\text{simp}}<10^{10}$ GeV the constraints on the elastic cross section on protons posed by Uranus's excess heat are stronger than the results of \cite{mcg}. Limits on baryonic dark matter candidates have been obtained in \cite{far,new}.

\section{Acknowledgments}
Many thanks go to Morris Podolak for going through the trouble of generating the temperature and density profiles for Uranus used in this article. It would have been impossible to accurately determine the exclusion limits for the S-wave annihilation cross section without this data. 
I thank Sabine Stanley for some useful references about Uranus, Benjamin Wandelt for references on recent work on constraining SIMPs, C\'eline Boehm for suggesting important improvements to an earlier draft version of this article and Gabrijela Zaharijas for making available the manuscript of \cite{new} prior to publication.

\appendix
\section{Calculation of the capture rate}
In this appendix we will derive Eq.\ \ref{rcapt} for the capture rate of light SIMPs. We will first derive an approximate analytical expression and then modify this somewhat to fit the results of simulations given in \cite{zah}. The analytical expression will be derived by estimating how many collisions are typically needed for a SIMP to slow down below escape speed. We then evaluate the probability that a SIMP entering the planet stays below the surface after this number of collisions.

We will assume that the mass of the SIMP, $m_{\text{simp}}$, is much less than the typical nucleon mass and that elastic cross sections with nuclei are isotropic. Under these conditions a SIMP inside a planet will perform an isotropic random walk. After each collision the fractional energy loss is a random number, uniformly distributed in the interval $I$:
\begin{equation}
I=\rhaak{0, 1 - \haak{\frac{1-r}{1+r}}^2}.
\end{equation}
Here $r=m_{\text{simp}}/m_{\text{nucleon}}$. This is true for general $r$, not just when $r<<1$. We can thus easily estimate how many scatterings are typically necessary to reduce the energy from $\half m_{\text{simp}} v^{2}$ to $\half m_{\text{simp}} \haak{v_{\text{esc}}}^2$, where $v_{\text{esc}}$ is the escape velocity from the planet. Denoting the ratio of the energies after and before the $j$-th scattering by $\gamma_{j}$, we can estimate the energy after $N$ scatterings, $E_{N}$, as follows. We first write:
\begin{equation}\label{en1}
E_{N}=  E_{0}\prod_{j=1}^{N}\gamma_{j}= E_{0}\exp\rhaak{\sum_{j=1}^{N}\log\haak{\gamma_{j}}}.
\end{equation}
Let's denote the relative probability that the SIMP will scatter off a nucleus of type $p$ by $f_{p}$. Typically, a fraction $f_{p}$ of the $\gamma_{j}$ in \eqref{en1} is  an energy ratio before and after scatterings off nuclei of type $p$. These $\gamma$'s are uniformly distributed in the interval $\rhaak{\haak{\frac{1-r_{p}}{1+r_{p}}}^2,1}$, where $r_{p}$ is the ratio of the mass of the SIMP and a nucleus of type $p$. The average of $\log\haak{\gamma}$ is thus:
\begin{equation}
\gem{\log\haak{\gamma}}=1+\frac{\haak{1-r_{p}}^{2}\log\lhaak{\frac{1-r_{p}}{1+r_{p}}}}{2 r_{p}}.
\end{equation}
By replacing the $\log\haak{\gamma}$'s by their averages in \eqref{en1} the following approximation is obtained:
\begin{equation}\label{en}
E_{N}\approx  E_{0}\exp\rhaak{-N\sum_{p} f_{p}\haak{1+\frac{\haak{1-r_{p}}^{2}\log\lhaak{\frac{1-r_{p}}{1+r_{p}}}}{2 r_{p}}}}.
\end{equation}
It follows from \eqref{en} that the number of collisions needed for a SIMP to be slowed down from a speed of $v$ to the escape velocity $v_{\text{esc}}$, $N\haak{v}$, is typically:
\begin{equation}\label{nv}
\begin{split}
N\haak{v} & =\frac{2\log\haak{\frac{v}{v_{\text{esc}}}}}
{\sum_{p}f_{p}\haak{ 1 +\frac{\haak{1-r_{p}}^{2}}{2 r_{p}}\log\lhaak{\frac{1-r_{p}}{1+r_{p}}}}}
\\ & \approx\frac{m_{\text{eff}}}{m_{\text{simp}}}\log\haak{\frac{v}{v_{\text{esc}}}}.
\end{split}
\end{equation}
Here we have used that $r_{p}<<1$ and defined
\begin{equation}
m_{\text{eff}}^{-1}\equiv \sum_{p}f_{p}m_{p}^{-1}
\end{equation}
where $m_{p}$ is the mass of a nucleus of type $p$.

Next we will evaluate the probability that a SIMP at some distance below the surface will remain below the surface after $N\haak{v}$ scatterings. As long as $\sigma n R >> 1$, we can treat this problem as a random walk in a half infinite space with absorbing boundary conditions. If $N\haak{v}$ is not too small, the random walk can be treated in the continuum limit where it becomes a diffusion problem. To find the continuum limit in this case, consider the evolution of the probability distribution of a SIMP in an infinite medium. The probability density $Q\haak{x,x\prime}$ that a SIMP will scatter at position $x$ if the previous scattering was at position $x\prime$ is: 
\begin{equation}\label{prop}
Q\haak{x,x\prime}=\frac{\sigma n\exp\haak{-n\sigma\lhaak{x-x\prime}}}{4\pi\haak{x-x\prime}^2}.
\end{equation}
The probability distribution for the SIMP after $N$ scatterings, $P_{N}\haak{x}$, evolves according to:
\begin{equation}\label{evol}
P_{N+1}\haak{x}=\int d^{3}x\prime P_{N}\haak{x\prime}Q\haak{x,x\prime}.
\end{equation}
It follows from \eqref{prop} and \eqref{evol} that the Fourier transform of $P_{N}\haak{x}$, defined as:
\begin{equation}
\hat{P}_{N}\haak{k}=\int d^3x P_{N}\haak{x} \exp\haak{-2\pi\imath k x}.
\end{equation}
satisfies
\begin{equation}
\hat{P}_{N+1}\haak{k}=\hat{P}_{1}\haak{k}\rhaak{\frac{n\sigma}{2\pi\lhaak{k}}\arctan\haak{\frac{2\pi\lhaak{k}}{n\sigma}}}^{N}.
\end{equation}
For large $N$ this becomes:
\begin{equation}
\hat{P}_{N+1}\haak{k}\approx\hat{P}_{1}\haak{k}\exp\rhaak{-\frac{N}{3}\haak{\frac{2\pi k}{n\sigma}}^{2}}.
\end{equation}
This implies:
\begin{equation}\label{evoleff}
P_{N+1}\haak{x}\approx\int d^{3}x\prime P_{1}\haak{x\prime}\tilde{Q}_{N}\haak{x,x\prime}
\end{equation}
where $\tilde{Q}_{N}\haak{x,x\prime}$ is:
\begin{equation}
\tilde{Q}_{N}\haak{x,x\prime}=\haak{\frac{3 n^2\sigma^2}{4\pi N}}^{\frac{3}{2}}\exp\haak{-\frac{3 n^2\sigma^2 \haak{x-x\prime}^2}{4 N}}.
\end{equation}

To find the evolution of the probability distribution of a SIMP in a half infinite space, we must put $P_{N}\haak{x}=0$ at the boundary. For problems that can be formulated entirely in the continuum limit this can be exactly achieved using the method of images, see e.g.\ \cite{img}. According to this method, if $g\haak{x}$ is the initial probability distribution of the SIMP defined in the half infinite space, we should substitute for $P_{1}\haak{x}$ in \eqref{evoleff}:
\begin{equation}\label{p1}
P_{1}\haak{x\prime}=g\haak{x\prime}-g\haak{R\haak{x\prime}}
\end{equation}
where $R\haak{x\prime}$ represents the vector obtained by reflecting $x\prime$ in the boundary of the half space. We expect that this method will yield a reasonable approximation. To obtain the number of SIMPs captured by a planet per unit time and per unit area we need to substitute for $g\haak{x\prime}$ the number density of SIMPs scattering for the first time at $x\prime$ per unit time and per unit volume in velocity space, $\phi_{1}$:
\begin{equation}\label{phi}
\phi_{1}\haak{a,v}=n_{\text{simp}}\exp\haak{-\frac{n\sigma a}{\cos\haak{\theta}}}n\sigma\tilde{f}\haak{v}\lhaak{v}.
\end{equation}
Here $a$ is the depth below the surface, $n_{\text{simp}}$ the halo number density of SIMPs and $\theta$ is the angle of the velocity w.r.t.\ the normal of the surface. The captured flux of SIMPs, $\mathcal{F}_{\text{captured}}$, can now be obtained by substituting \eqref{phi} in \eqref{p1} and the resulting expression for $P_{1}$ in \eqref{evoleff}. By integrating this over $x\prime$ and $x$ over the half infinite space below the surface we obtain to lowest order in $1/N\haak{v}$:
\begin{equation}\label{fiv}
\phi\haak{v}\approx\frac{2}{\sqrt{3 \pi}}n_{\text{simp}}N\haak{v}^{-\frac{1}{2}}\pi \tilde{f}\haak{v}v^3.
\end{equation}
Here $\phi\haak{v} dv$ is the flux of captured particles with original speeds between $v$ and $v+dv$. For speeds $v \sim v_{\text{esc}}$ the above derivation is not very accurate because then $N\haak{v}$ is not large. However, since $\tilde{f}\haak{v}$ is small in this case, this isn't a problem. We shall neglect the flux of particles moving slower than $v_{\text{esc}}$. Integrating \eqref{fiv} and using Eq.\ \ref{nv} for $N\haak{v}$, we obtain for the flux of captured particles:
\begin{equation}\label{capt1}
\mathcal{F}_{\text{captured}}=n_{\text{simp}}\int_{v_{\text{esc}}}^{\infty}R\haak{v}\pi\tilde{f}\haak{v}v^3 dv.
\end{equation}
Here
\begin{equation}\label{rvcp}
R\haak{v}\approx\frac{2}{\sqrt{3 \pi}}\sqrt{\frac{m_{\text{simp}}}{m_{\text{eff}}}}\haak{\log\haak{\frac{v}{v_{\text{esc}}}}}^{-\frac{1}{2}}
\end{equation}
is the probability that a particle moving with speed $v$ into the planet is captured.

In \cite{zah} the capture of strongly interacting dark matter particles heavier than 2 GeV by the Earth is investigated using simulations. We find that Eq.\ \ref{rvcp} predicts a captured flux that is about a factor of 2 too low. The likely source of this difference is the improper use of the method of images to impose absorbing boundary conditions. Changing the prefactor of $\frac{2}{\sqrt{3 \pi}}$ to $1.27$ gives an almost perfect fit to the captured fraction of the flux in the mass range 2 GeV-10 GeV. For Earth, $m_{\text{eff}}\approx 21.5$ GeV, about ten times as high as for Uranus. When $m_{\text{simp}}$ is below 10 GeV, the ratio $m_{\text{simp}}/m_{\text{eff}}$ for Earth is thus in roughly the same range as the ratio $m_{\text{simp}}/m_{\text{eff}}$ for Uranus when $m_{\text{simp}}$ is below 1 GeV. We thus expect that 
\begin{equation}\label{capt}
R\haak{v}\approx 1.27 \sqrt{\frac{m_{\text{simp}}}{m_{\text{eff}}}}\haak{\log\haak{\frac{v}{v_{\text{esc}}}}}^{-\frac{1}{2}}
\end{equation}
is a reasonable approximation for the capture probability.

\end{document}